\documentstyle[preprint,aps]{revtex}
\def\bra#1{\langle #1 \,\vert}
\def\ket#1{\vert\, #1 \rangle}

\def\ra {\rightarrow}
\def\lag{\langle}
\def\rag{\rangle}
\def\Tr{{\rm Tr}}

\def\MeV{\mbox{MeV}}
\def\fm{\mbox{fm}}

\def\MeV{\mbox{MeV}}
\def\d{\mbox{d}}
\def\e{\mbox{e}}
\def\sign{\mbox{sign}}
\def\fm{\mbox{fm}}
\begin{document}
\title{$E2/M1$ Ratio for the $\gamma N\rightarrow \Delta$ Transition
in the Chiral Quark Soliton Model}
\author{T. Watabe\thanks{watabe@baryon.tp2.ruhr-uni-bochum.de},
Chr.V. Christov\thanks{christov@neutron.tp2.ruhr-uni-bochum.de}
and K. Goeke\thanks{goeke@hadron.tp2.ruhr-uni-bochum.de}}
\address{ Institut f\"{u}r Theoretische Physik II, Ruhr-Universit\"at
Bochum, D-44780 Bochum, Federal Republic Germany}
\maketitle
%
\tighten
\begin{abstract}
We calculate the electric quadrupole to magnetic dipole transition
ratio $E2/M1$ for the reaction $\gamma N \ra \Delta(1232)$ in the
chiral quark soliton model. The calculated $E2/M1$ ratio is in a good
agreement with the very new experimental data.
We obtain non-zero negative value for the electric quadrupole $N-\Delta$
transition moment, which suggests an oblate deformed charge structure
of the nucleon or/and the delta isobar. Other observables related to this
quantity, namely the $N-\Delta$ mass splitting, the isovector charge
radius, and isovector magnetic moment, are properly reproduced as well.
\end{abstract}
\draft
\newpage
%
The ratio of electric quadrupole to magnetic dipole amplitude
($E2/M1$) for the reaction $\gamma + N \ra \Delta(1232)$ is a quantity
sensitive to a presence of charge deformations in the baryon structure.
The most reliable phenomenological estimate $E2/M1 = (-1.5 \pm
0.2)\%$ so far comes from detailed analysis~\cite{Davidson86} of the
available photoproduction
data, assuming the most general $\gamma N \Delta$ gauge coupling and
taking into account the unitary condition via Watson theorem. Very
recently a very precise $\pi^{0(+)}$-photoproduction
experiment~\cite{Beck95} has been performed at
MAMI, Mainz, which allows for a direct model-independent estimate of the
ratio $E2/M1$. The preliminary result\footnote{In fact, the authors
have faced this experimental result after completing the calculations.}
 $E2/M1 = (-2.4 \pm 0.2)\%$~\cite{Beck95} confirms the negative sign
suggested from the analysis~\cite{Davidson86} and shows a larger
$E2/M1$-asymmetry. This non-zero negative value
is a clear indication for the presence of an oblate charge deformation in
the nucleon or/and delta and as such it imposes
strong constraints for the effective models of baryon structure.  Wirzba
and Weise have investigated the $E2/M1$ ratio in a modified Skyrme model
which includes stabilizing fourth- and sixth-order terms~\cite{Wirzba87}.
They obtained values between $-5\%$ and $-2.5\%$ depending on the
coupling parameters of stabilizing terms. However, the other related
observables, namely the $N-\Delta$ mass difference, charge radii, and
the isovector magnetic moment, are not properly described.

In the present work we study $E2/M1$ ratio for the process $\gamma + N \ra
\Delta(1232)$  in the chiral quark soliton model (for review see
ref.~\cite{Meissner94}). We employ the simplest SU(2)-version
of the model with up and down quark degenerated in mass, which is
based on the semibosonized Nambu Jona-Lasinio
lagrangean~\cite{Nambu61,Eguchi76}:
\begin{equation}
{\cal L} = \bar{\Psi}(-i \gamma^{\mu} \partial_{\mu}
+ m_0 + M U^{\gamma_5}) \Psi,
\label{lagrangean}
\end{equation}
with auxiliary meson fields
\begin{equation}
U(\vec x) = \e^{i \vec{\tau} \cdot \vec{\pi}(\vec x) / f_{\pi}}\,,
\label{u}
\end{equation}
constrained on the chiral circle. The model is non-renormalizable and a
finite cutoff is needed. The latter is treated as a parameter of the model
and together with the current quark mass $m_0$ is fixed in the
mesonic sector to reproduce the physical pion mass $m_\pi$ and the pion
decay constant $f_\pi$. The last model parameter, the constituent quark mass
$M$, can be related to the empirical value of the quark condensate
but actually it still leaves a broad range for $M$.\footnote{Actually, in
order to obtain a good description of the baryonic properties the mass $M$
has to be chosen around $420$ MeV (see review~\cite{Meissner94}).}

In the model the baryons appear as a bound state of $N_c$ (number of colors)
valence quarks coupled to the polarized Dirac sea. Since the model lacks
confinement the proper way to describe the nucleon is to
consider~\cite{Diakonov88} a correlation function of two $N_c$-quark
currents with nucleon quantum numbers $J J_3, T T_3$ at
large euclidean time-separation:
\begin{eqnarray}
\lim\limits_{T\to \infty}\Pi_N(T)&=&\lag J_N(\vec x,+T/2)
J_N^\dagger(\vec y,-T/2) \rag\nonumber \\
&=&{1\over Z}\,\Gamma^{\{f\}}_{JJ_3,TT_3}\,\Gamma^{\{g\}*}_{JJ_3,TT_3}\,
\int{\cal D}U
\prod\limits_{i=1}^{N_c}\lag{T/2,{\vec x}}|\frac
1D|-T/2,{\vec y}\rag_{f_i g_i}\e^{N_c\Tr\log D(U)}\,,
\label{correlator} \end{eqnarray}
where current $J_N$ is a composite quark operator~\cite{Ioffe81}:
\begin{equation}
J_N(\vec x,t) = \frac{1}{N_c !} \
\varepsilon^{\beta_1 \cdots \beta_{N_c}} \
\Gamma^{ \{ f\} }_{J J_3, T T_3} \
\Psi_{\beta_1 f_1}(\vec x,t) \cdots \Psi_{\beta_{N_c} f_{N_c}}(\vec
x,t)\,.
\label{jn}
\end{equation}
In the above path integral the quark fields are integrated out. The
effective action includes the Dirac operator
\begin{equation}
D(U)=i\partial_t-h(U),
\label{DIRAC}\end{equation}
with one-particle hamiltonian given by
\begin{equation}
h(U)=\frac {\vec{\alpha}\cdot\vec{\nabla}}i+\beta M U^{\gamma_5}+\beta
m_0\,.
\label{HAMIL}\end{equation}

In the model the nucleonic solution is obtained in two steps. In the
first step, in leading order in $N_c$ the integral over the meson fields $U$
in eq.(\ref{correlator}) is done in a saddle point approximation.
To that end we look for a stationary localized meson configuration (soliton)
of hedgehog structure
\begin{equation}
\bar U(x)\,=\,\e^{i\vec{\tau}\cdot\hat x\,P(x)}\, ,
\label{HEDGEHOG}\end{equation}
which minimizes the effective action. Actually,
the soliton solution is found  by solving the corresponding equations of
motion in an iterative self-consistent procedure~\cite{Meissner94}.
Since the hedgehog soliton $\bar U(x)$ does not preserve the spin and
isospin, as a next step we make use of the rotational zero modes to
quantize it. It is done assuming a rotating meson hedgehog
fields of the form
\begin{equation}
U({\vec x},t)=R(t)\,\bar U({\vec x})\, R^+(t)\,
\label{ROTHEDG}\end{equation}
with $R(t)$ being a time-dependent rotation SU(2) matrix in the isospin
space. It is easy to see that for such an ansatz one can
transform the effective action
\begin{equation}
\Tr\log D(U)=\Tr\log(D(\bar U)-\Omega)
\label{ROTEFA}\end{equation}
in order to separate the angular velocity matrix:
\begin{equation}
\Omega=-iR^+(t)\dot R(t)=\frac 12\Omega_a\tau_a\,.
\label{OMEGA}\end{equation}
Since the angular velocity is quantized according to the canonical
quantization rule, it appears as $\Omega_a\sim \frac 1{N_c}$. This allows
one to consider
$\Omega$ as perturbation and to evaluate any observable as a perturbation
series in $\Omega$ which is actually an expansion in $\frac 1{N_c}$.
The path integral over $R(t)$, which appears due to the ansatz
(\ref{ROTHEDG}), determines the spin-flavor structure of the nucleonic
solution. For given spin $J,J_3$ and isospin $T,T_3$ this structure
can be expressed through the Wigner $D$ function
\begin{equation}
|N,T_3J_3 \rangle (R) = (-1)^{T+T_3}\sqrt{2T+1}\,D^{T=J}_{-T_3,J_3}(R)\,.
\label{D-functions}
\end{equation}

Very recently, the nucleon electromagnetic form factors have been
calculated~\cite{Christov95} in this semiclassical quantization scheme up to
the next to leading order in angular velocity.  The results for the
constituent quark mass $M\approx 420$ MeV are in fairly
good agreement with the experimental data. It should be also noted
that the $1/N_c$ rotational corrections improve considerably the
theoretical value for the isovector magnetic moment~\cite{Christov94}.

Obviously it is worthily
to consider in this model also the ratio $E2/M1$ for the process $\gamma N
\longrightarrow \Delta$ whose value is well settled by the recent
experiment in Mainz~\cite{Beck95}. It is defined~\cite{Tanabe76} as a
ratio
\begin{equation}
\frac{E2}{M1} \equiv \frac{1}{3} \
\frac{M^{E2}(\vec{k}, \lambda = +1 \ ; \
p(J_3 = -\frac{1}{2}) \ra \Delta^+(J_3 = +\frac{1}{2}))}
{M^{M1}(\vec{k}, \lambda = +1 \ ; \
p(J_3 = -\frac{1}{2}) \ra \Delta^+(J_3 = +\frac{1}{2}))}
\label{deftr}\end{equation}
of electric quadrupole amplitude $M^{E2}$ to magnetic dipole one
$M^{M1}$.
Here $k$ is the momentum  of a photon of helicity $\lambda$
in the $\Delta$ rest frame:
\begin{equation}
k = \frac{M_\Delta^2 - M_N^2}{2 M_\Delta}\,.
\label{restk}
\end{equation}
Both amplitudes are related to the corresponding matrix element of the
isovector current:
\begin{equation}
j^\mu_3(\vec{x}) = \bar{\Psi}(\vec{x})
\gamma^\mu {\tau_3\over 2} \Psi(\vec{x}) .
\label{current}
\end{equation}
For the electric quadrupole amplitude $M^{E2}$, according to the
Siegert's theorem~(see for instance~\cite{Greiner70}), one can use the zero
component $j^0_a$ as well the space component -- $\nabla\cdot\vec j_a$.
Similar to ref.~\cite{Wirzba87} we decide to express the amplitude $M^{E2}$
via the $N-\Delta$ transition matrix element of $j_3^0$ (charge
density) of the current $j^\mu$:
\begin{equation}
M^{E2}(\vec{k}, \lambda ; N \ra \Delta) =
\sqrt{15 \pi} \int d^3x \ \lag \Delta \vert j^0_3(\vec{x}) |
N \rag Y_{2 \lambda} (\hat{x}) j_2(kx) ,
\label{mel}
\end{equation}
The reason is that in the present model this quantity can be
calculated~\cite{Christov95} directly in terms of quark
matrix elements whereas the for $\nabla\cdot\vec j_a$ one should use
supplementary the classical equations of motion (saddle point), for which
the ansatz (\ref{ROTHEDG}) is apparently not a solution.

The amplitude $M^{M1}$ is directly related to the $N-\Delta$ transition
matrix element of the space components $j_a^k$:
\begin{equation}
M^{M1}(\vec{k}, \lambda ; N \ra \Delta) =
- \lambda {3\over 2} \int d^3x \lag \Delta | (\hat x \times
\vec j_3)_\lambda| N \rag j_1(kx) .
\label{mml}
\end{equation}
In the case of $\lambda = +1$,  $J_3^p = -\frac{1}{2}$
$J_3^\Delta = +\frac{1}{2}$,
we have
\begin{equation}
M^{E2} = +\frac{15 \sqrt{3}}{4} \int\d r r^2 j_2(kr)\rho^{E2}_{N
\Delta}(r)\,,
\label{me2}
\end{equation}
and
\begin{equation}
M^{M1}= - 3 \int \d r r^2 j_1(kr) \rho^{M1}_{ N \Delta}(r)\,,
\label{mm1}
\end{equation}
respectively.

For the matrix element of the isovector current $j^\mu_3$ we follow the line
of ref.~\cite{Christov95} and  in fact, we use the results presented there.
Here we will only sketch the derivation. We evaluate the expectation value
of the quark bilinear operators,
$\Psi^\dagger \gamma^0\gamma^\mu {\tau_3\over 2} \Psi$,
represented by the euclidean functional integral~\cite{Diakonov88} with
lagrangean (\ref{lagrangean}):
\begin{eqnarray}
&&\langle N', {\vec p}' | \Psi^\dagger(0)\gamma^0
\gamma^\mu {\tau_3\over 2}
  \Psi(0) | N, {\vec p} \rangle
=\lim_{\scriptstyle T^\prime\to+\infty \atop \scriptstyle T\to -\infty}
\frac 1Z \int \d^3 x \d^3y
\e^{p_4^\prime T^\prime - p_4 T - i{\vec p}' {\vec x}' + i{\vec p} {\vec
x}}\nonumber \\
&&\times\int{\cal D}U \int {\cal D}\Psi \int {\cal D}\Psi^\dagger
J_{N^\prime}(\vec x^\prime,T^\prime) \> \Psi^\dagger(0) \gamma^0\gamma^\mu
{\tau_3\over 2} \Psi(0) J_N^\dagger(\vec x,T) \e^{- \int d^4 z
\Psi^\dagger D(U) \Psi }\,.
\label{form-factor-integral}
\end{eqnarray}
Integrating out the quarks in (\ref{form-factor-integral}) it is easy to
see
that the result is naturally split in valence and sea parts.  After that
we follow the same steps as for correlation
function (\ref{correlator}). First, we integrate over the meson fields in
saddle point approximation (large $N_c$ limit). As a second step, we take
into account the rotational zero modes using the ansatz (\ref{ROTHEDG}).
Due to the collective path integral over $R(t)$ our scheme\footnote{
The details of this procedure can be found in ref.\cite{Christov95}.}
involves a time-ordered product of the collective operators
\begin{equation}
\Omega_a(R(t))=-i\Tr \bigl(R^\dagger(t)\dot R(t)\tau_a\bigr)\quad
\mbox{and}\quad
D_{ab}(R(t))= {1\over 2}\Tr \bigl(R^\dagger(t)\tau_a R(t)\tau_b\bigr)\,,
\label{colloper}
\end{equation}
which appear after the expansion in $\Omega$.

In the above scheme the matrix element of the time-component of the
isovector current,
calculated in the semiclassical quantization scheme, includes only terms
linear in $\Omega\sim 1/N_c$
\begin{eqnarray}
\lag \Delta | j_0^3(\vec x) | N \rag  &=&  {N_c\over 2 \Theta} \,
 \biggl\{\sum\limits_{m,n}
 {\cal R}_\Theta^\Lambda(\epsilon_m,\epsilon_n)
\biggl(\Phi_m^\dagger({\vec x}) \, \tau_a  \, \Phi_n({\vec x})\biggr) \,
\langle n |\tau_c| m \rangle \nonumber\\
&-&\sum\limits_{n\neq {val}}
\frac{1}{\epsilon_{val} - \epsilon_n}\biggl(\Phi_n^\dagger({\vec x})\tau_a
\Phi_{val}({\vec x})\biggr) \,\langle val |\tau_c| n \rangle\biggl\}
\langle \Delta, J_3^\prime T_3^\prime|\{J_c,D_{3a}\}|N, J_3 T_3
\rangle\,, \label{electric-isovector-form-factor} \end{eqnarray}
whereas the matrix element of the space-components of the isovector
current includes leading order terms $\sim \Omega^0$ as well as
next to leading order ones $\sim\Omega$ ($1/N_c$):
\begin{eqnarray}
&&\bra{\Delta} j^k_3(\vec x)\ket{N}=
N_c\Biggl\{\Bigl(\Phi^\dagger_{val}({\vec
x})\,\gamma^0\gamma^k\tau_a\,\Phi_{val}({\vec
x})\Bigr)\bra{\Delta,J_3^\prime
T_3^\prime}D_{3b}\ket{N,J_3T_3}\nonumber\\
&&+\frac i{2\Theta}\sum\limits_{n\neq val}\sign({\epsilon_n})
\frac {\Bigl(\Phi^\dagger_{val}({\vec
x})\gamma^0\gamma^k\tau_b\Phi_n({\vec
x})\Bigr)\bra{n}\tau_c\ket{val}}{\epsilon_n -
\epsilon_{val}}\bra{\Delta,J_3^\prime T_3^\prime}[\hat J_c
\,,\,D_{3b}]\ket{N,J_3T_3}\nonumber\\
&&-\sum\limits_n {\cal R}_{M1}^\Lambda(\epsilon_n)
\Bigl(\Phi^\dagger_n({\vec x}) \gamma^0 \gamma^k \tau_b
\Phi_n({\vec x})\Bigr)\bra{\Delta,J_3^\prime
T_3^\prime}D_{3b}\ket{N,J_3T_3}
\nonumber\\
&&+{i\over \Theta}  \sum\limits_{n,m}{\cal
R}_{M2}^\Lambda(\epsilon_m,\epsilon_n)\Bigl(\Phi_m^{\dagger}({\vec x})
\gamma^0 \gamma^k \tau_b\Phi_n({\vec x})\Bigr)
\bra{n}\tau_c\ket{m}\bra{\Delta,J_3^\prime T_3^\prime}[\hat J_c\,
,\,D_{3b}]\ket{N,J_3T_3}\Biggr\}\,. \label{MVFDS2a}\end{eqnarray}
Here $\Theta$ is the moment of inertia, and $\Phi_n$ and $\epsilon_n$ are
the eigenfunctions and the eigenvalues of the hamiltonian (\ref{HAMIL}). The
regularization functions ${\cal R}_\Theta^\Lambda$, ${\cal R}_{M1}^\Lambda$
and ${\cal R}^\Lambda_{M2}$ can be found in ref.\cite{Christov95}.

For completeness we present the final results for the electric quadrupole
density $\rho^{E2}_{N \Delta}(r)$, split in valence and Dirac sea parts,
\begin{equation}
\rho^{E2 ; val}_{N \Delta}(r) =
\frac{N_c}{\Theta} \frac{\sqrt{6\pi}}{90}
\sum_{n \neq val} \frac{1}{\epsilon_n - \epsilon_{val}} \
\biggl(\Phi_{val}(r)\| [Y_2 \otimes \tau^{(1)}]^{(1)}\|
\Phi_n(r)\biggr) \lag val \| \tau^{(1)} \| n \rag\,,
\label{re2val}
\end{equation}
and
\begin{equation}
\rho^{E2 ; sea}_{N \Delta}(r) =
\frac{N_c}{\Theta} \frac{\sqrt{6\pi}}{180}
\sum_{n,m = all} R_\Theta^\Lambda(\epsilon_n,\epsilon_m)
\biggl(\Phi_n(r)\|[Y_2 \otimes \tau^{(1)}]^{(1)}\|
\Phi_m(r)\biggr) \lag n \| \tau^{(1)} \| m \rag
\label{re2vp}
\end{equation}
as well as for the magnetic density $\rho^{M1}_{N\Delta}(x)$:
\begin{equation}
\rho^{M1 ; val(\Omega^0)}_{N \Delta}(r) =
- N_c \frac{i}{6 \sqrt{6}} \biggl(\Phi_{val}(r)
\|\gamma_5[[\hat{x}^{(1)}\otimes\sigma^{(1)}]^{(1)}\otimes\tau^{(1)}]^{(0)}\|
\Phi_{val}(r)\biggl)\,, \label{rm1val0}
\end{equation}
\begin{equation}
\rho^{M1 ; sea(\Omega^0)}_{N \Delta}(r) =
- N_c \frac{i}{12 \sqrt{6}}
\sum_{n = all} R_{M1}^\Lambda(\epsilon_n)\sqrt{2 K_n + 1}
\biggl(\Phi_n(r)
\|\gamma_5[[\hat{r}^{(1)}\otimes^{(1)}\sigma]^{(1)}\otimes\tau^{(1)}]^{(0)}\|
\Phi_n(r)\biggr) \,, \label{rm1vp0}
\end{equation}
\begin{equation}
\rho^{M1 ; val(\Omega^1)}_{N \Delta}(r) =
- \frac{N_c}{\Theta} \frac{i}{36} \sum_{n \neq val}
\frac{\sign(\epsilon_n)}{\epsilon_n-\epsilon_{val}} \biggl(\Phi_{val}(r)
\|\gamma_5[[\hat{r}^{(1)}\otimes\sigma^{(1)}]^{(1)}\otimes
\tau^{(1)}]^{(1)}\|
\Phi_n(r)\biggr) \lag val \| \tau^{(1)} \| n \rag ,
\label{rm1val1}
\end{equation}
\begin{eqnarray}
\rho^{M1 ; sea(\Omega^1)}_{N \Delta}(r) =
- \frac{N_c}{\Theta} \frac{i}{72} \sum_{n,m = all}&&
R_{M2}^\Lambda(\epsilon_n-\epsilon_m) \biggl(\Phi_n(r)
\|\gamma_5[[\hat{r}^{(1)}\times\sigma^{(1)}]^{(1)}\times\tau^{(1)}]^{(1)}\|
\Phi_m(r)\biggr) \nonumber\\
&&\times\lag n \| \tau^{(1)} \| m \rag \,.
\label{rm1vp1}
\end{eqnarray}
On the other hand, using (\ref{mel}) and (\ref{mml}) in the
approximation $k \cdot R \ll 1$, where $R$ is the nucleon charge radius, we
get following simple formulae
\begin{equation}
M^{E2}=- \frac{3}{4\sqrt{2}} k^2 \lag Q_{zz} \rag_{N \Delta} ,
\label{sime2}
\end{equation}
\begin{equation}
M^{M1}=- \frac k{\sqrt{2}} \frac{1}{2 M_N} \ \mu_{N\Delta},
\label{simm1}
\end{equation}
where $\lag Q_{zz} \rag_{N\Delta}$ is the electric quadrupole
transition moment and $\mu_{N\Delta}$ is the transition magnetic moment:
\begin{equation}
\mu_{N\Delta}= \frac{1}{\sqrt{2}} \mu_{I=1}\,.
\label{trnfm}
\end{equation}
Here $\mu_{I=1}=\mu_p-\mu_n$ is the isovector magnetic moment.
Using (\ref{sime2}) and (\ref{simm1}), in the $k \cdot R \ll 1$
approximation one can relate the ratio $E2/M1$ to the electric
quadrupole $N\Delta$ transition moment:
\begin{equation}
\frac{E2}{M1} = \frac{1}{2} k M_N
\frac{\lag Q_{zz} \rag_{N\Delta}}{\mu_{N\Delta}}\,.
\label{rtlw}
\end{equation}
It should be noted that in contrast to the Skyrme model, ${\lag Q_{zz}
\rag}_{N\Delta}$ cannot be directly related to the isovector charge
radius.
The corresponding transition charge density $\rho^{E2}_{N \Delta}(x)$ has a
more complicated structure, including a spherical harmonics tensor $Y_{2
\mu}$, which acts on the quark wave function as a projector for the charge
deformation.

In the numerical computations we use the method of Ripka and
Kahana~\cite{Ripka84} for solving the eigenvalue problem in a finite
quasi--discrete basis.

In table~\ref{Tabl1}, we present our results for the ratio $E2/M1$
as well as for some related observables, namely
the isovector charge m.s.radius, the $N-\Delta$ transition magnetic moment
$\mu_{N\Delta}$, the $N-\Delta$ mass difference, and the quadrupole
electric transition moment $\lag Q_{zz}\rag_{N\Delta}$, for three different
values of the constituent quark mass $M$. We compare our results
with the experiment as well as with the numbers of
ref.~\cite{Wirzba87}. With constituent quark mass $M$ around 420 MeV we
obtain for the $E2/M1$ ratio values between $-2.5\%$ and $-2.3\%$  quite in
agreement with the last experiment data~\cite{Beck95}. It should be
mentioned that for the same values of the constituent quark mass $M$
the nucleon properties (including also the nucleon form factors)
are reproduced fairly well~\cite{Christov95}. The only exception is the
$N-\Delta$ transition magnetic moment which is underestimated by 25\%.
Our results for other observables in table~\ref{Tabl1} show an overall
good agreement with the experiment which, however, is not the case for the
Skyrme model calculations~\cite{Wirzba87}. The Skyrme model results show
a strong underestimation of the isovector charge radius and $N-\Delta$
mass splitting whereas the isovector magnetic moment is strongly
overestimated.

In the table~\ref{Tabl1} we also present the results for  the
ratio $E2/M1$ in the $k \cdot R \ll 1$ approximation. Despite that
this approximation is not justified it seems that it works in practice
satisfactorily: the numbers using the formula (\ref{rtlw}) overestimate the
exact results by not more than 20\%. From the relation (\ref{rtlw}) we
also get a rough estimate for the electric quadrupole transition moment
$\lag Q_{zz} \rag_{N\Delta}=-0.026$ using the experimental values for
$E2/M1=-2.4\pm 0.2$~\cite{Beck95} and $\mu_{N\Delta}=3.3$, which
is not far from our model prediction $-0.02$. This negative non-zero
value indicates a presence of an oblate type of charge deformations in the
nucleon or/and  delta structure. It is interesting to mention that the
dominant contribution to the  $\lag Q_{zz} \rag_{N\Delta}$ in the NJL
model comes from the Dirac sea. It can be seen  in the
table~\ref{Tabl1} as well on the figure~\ref{Figr1}, where the electric
quadrupole transition moment density, separated in valence and Dirac
sea parts, is shown. It means that the main charge deformation is due
to the polarized Dirac sea, whereas the
valence quarks are almost spherically distributed. Since using the
gradient expansion, the polarization of the Dirac sea can be expressed
in terms of the dynamical pion field -- pion cloud, one can think of
the nucleon or/and of the delta as consisting of an almost spherical
valence quark core surrounded by a deformed pion cloud.

In summary, we study the electric quadrupole to magnetic dipole
transition ratio $E2/M1$  for the reaction $\gamma N \ra \Delta(1232)$
in the chiral quark soliton model. The calculated $E2/M1$ ratio for the
constituent mass around 420 MeV is in a good agreement with the very
new experimental data. We obtain a non-zero negative value for the
electric quadrupole transition moment, which suggests an oblate deformed
charge structure of the nucleon or/and for the delta isobar. Other
related observables, namely the $N-\Delta$ mass difference, the
isovector charge radius, and the $N-\Delta$ transition magnetic moment,
are properly reproduced as well.

\section*{Acknowledgement}

We would like to thank Reinhard Beck for making kindly the preliminary
experimental results and some details, concerning the experiment,
available to us prior the publication. The project has been partially
supported by the VW Stiftung, DFG and COSY (J\"ulich).

%
\begin{figure}
\caption{Electric quadrupole $N-\Delta$ transition moment densitiy,
separated in valence and Dirac sea parts.}
\label{Figr1}
\end{figure}
%
\begin{table}
\caption{Ratio $E2/M1$ and some related observables, calculated in
the NJL model for three different values of the constituent
mass$M=400, 420$ and $450$ MeV, compared with experimental values. The
Skyrme model results~\protect\cite{Wirzba87} are also presented.}
\label{Tabl1}
\begin{tabular}{ccccccccc}
 & \multicolumn{6}{c}{{\bf Constituent quark mass $M$}} & & \\
{\bf Quantity} &
\multicolumn{2}{c}{400 MeV} &
\multicolumn{2}{c}{420 MeV} &
\multicolumn{2}{c}{450 MeV} &
{\bf Skyrme} & {\bf Exper.} \\
\cline{2-7}
 & total & sea & total & sea & total & sea &  & \\
\hline
$\lag r^2 \rag_{I=1}  \ [\fm^2]$ &
0.88 & 0.35 & 0.84 & 0.37 & 0.79 & 0.41 & 0.60 & 0.86 \\
$\mu_{\Delta N}  \ [n.m.]$ &
2.34 & 0.57 & 2.28 & 0.58 & 2.20 & 0.60 & 6.18 & 3.33 \\
$M_\Delta - M_N  \ [\MeV]$ &
255  &      & 278  &       & 311  &      & 199  & 294  \\
$\lag Q_{zz} \rag_{\Delta N}$ \ [fm$^2$] &
$-$0.020 & $-$0.014 & $-$0.020 & $-$0.015 & $-$0.021 & $-$0.016 &
$-$0.028&$-0.026$\footnote{Using eq.(\ref{rtlw})}\\
$M^{E2}$ &0.012 & 0.008 & 0.013 & 0.008 & 0.013 & 0.009 & & \\
$M^{M1}$&$-$0.189 & $-$0.041 & $-$0.186 & $-$0.042 & $-$0.182 & $-$0.043&&\\
$E2/M1$\ [\%]& $-$2.19 & & $-$2.28& &$-$2.42& &$-2.6$&$-2.4\pm$0.2\\
$E2/M1$ \ [\%]$^4$&$-$2.64 & & $-$2.79 & & $-$2.99 & & & $-2.4\pm$0.2\\
\end{tabular}
\end{table}
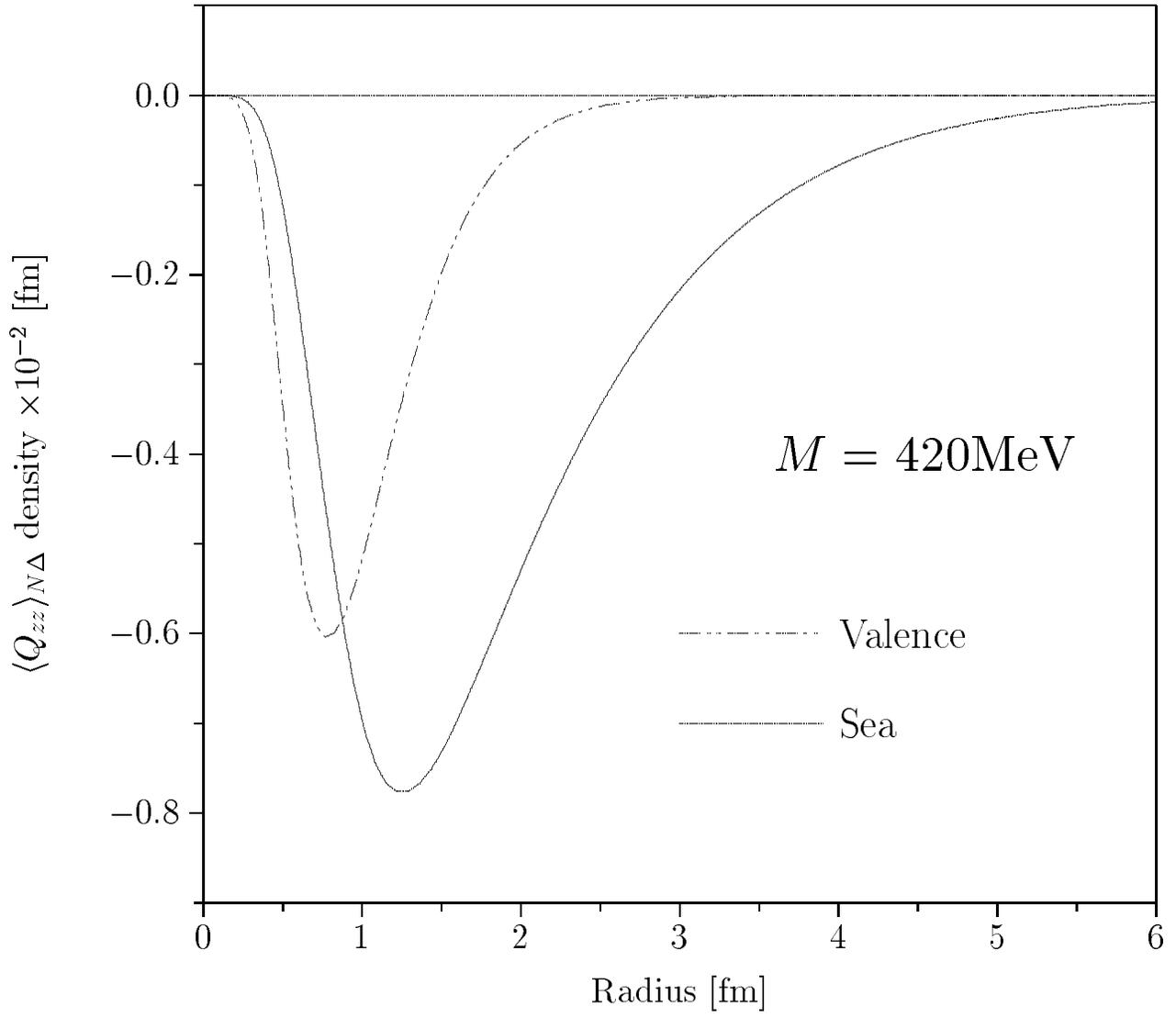
\end{document}